\documentclass[letterpaper,twocolumn,showpacs,pra,aps]{revtex4-1}

\usepackage{amsmath}  
\usepackage{amsfonts} 
\usepackage{graphicx} 
\usepackage{amssymb}

\begin{document}

\title{Solitons in the Einstein universe}

\author{Eugene B. Kolomeisky and Ephraiem Sarabamoun}

\affiliation
{Department of Physics, University of Virginia, P. O. Box 400714, Charlottesville, Virginia 22904-4714, USA}

\date{\today}

\begin{abstract}
We show that equations of Newtonian hydrodynamics and gravity with Einstein's cosmological constant included admit gravitostatic wave solutions propagating in the background of Einstein's static Universe.  In the zero pressure limit these waves exist at an average matter density exceeding that of Einstein's Universe.  They have the form of a lattice of integrable density singularities localized at the maxima of the gravitational potential.  These singularities are steady-state counterparts of the so-called Zeldovich pancakes (ZP), interim wall-like structures appearing at nonlinear stages of development of gravitational instability.  As the average matter density decreases, the period of the ZP lattice increases diverging at the density of Einstein's Universe.  Solitary wave solutions are found at exactly the density of Einstein's Universe, and at a slightly larger density the wave may be viewed as a lattice of well-separated ZP solitons.             
\end{abstract}


\maketitle

\section{Introduction}

It is well-known that in order to conform with astronomical observations of the day and his own views that the Universe is static, in 1917 Einstein modified \cite{Einstein} his equations of the general theory of relativity away from their original form \cite{LL2}.  The modification known as the cosmological or $\Lambda$-term has the effect of counteracting the long-range attractive gravitational interaction making a static Universe possible.  In 1922-1924, however, Friedmann \cite{Friedmann}  demonstrated that not only do Einstein's field equations have spatially homogeneous and isotropic solutions with or without the $\Lambda$-term, they are generally non-stationary, and Einstein's static Universe is a special degenerate case.  Results similar to Friedmann's have been also found in 1927 by Lema\^{\i}tre \cite{Le} who, based on astronomical observations, concluded that the Universe is expanding.  This picture of the Universe has been put on solid observational footing by Hubble's discovery of a linear velocity-distance relationship for distant galaxies \cite{Hubble}.  Moreover, Einstein's static Universe has been shown to be unstable with respect to small density perturbations \cite{Eddington}.  Eventually a consensus was reached that the introduction of the $\Lambda$-term was an unnecessary complication of the theory not supported by observational data of the day \cite{LL2}.   

The attitude toward the cosmological term started to shift about two decades ago when accelerated expansion of the Universe was discovered \cite{accelerated1,accelerated2}, and presently a case can be made in favor of a positive cosmological $\Lambda$-term as a possible explanation of the observations \cite{SS}.   Recent measurements of the Cosmic Microwave Background (CMB) \cite{CMB} have additionally supported the case in favor of constant $\Lambda>0$.  The latest measurements on bright stars in galaxies \cite{bright} however seem to indicate that $\Lambda$ has varied from the time the CMB was created to the present era.  Despite these differences that clearly will have to be addressed in future observations, there is solid observational evidence for the existence of the cosmological term.  The current value of the cosmological constant is \cite{CMB}
\begin{equation}
\label{Lambda}
\Lambda = (1.106 \pm 0.023) \times 10^{-56} cm^{-2}.
\end{equation}     

These developments renewed interest in cosmological models with the cosmological constant.  The goal of this paper is to provide a solution to one of such models.  Namely, we demonstrate the existence of traveling wave cosmological solutions that can propagate in the background of Einstein's static Universe.  These solutions also encompass solitary waves whose existence is only possible due to the presence of the cosmological constant.  

Hereafter our tool is Newtonian cosmology of Milne and McCrea \cite{Milne}. Compared to the general theory of relativity this approach offers simplicity and ease of extension to models lacking the homogeneity and isotropy of the classic Friedmann-Lema\^{\i}tre solutions.  While not a substitute for the general theory of relativity, in questions of cosmology, the Newtonian approach is nearly as rigorous as that of the general theory of relativity \cite{ZN}.  Indeed, the Friedmann-Lema\^{\i}tre dynamics is recovered within the Newtonian cosmology \cite{Milne}, and Newtonian analysis of gravitational instability \cite{Bonnor} agrees with that based on the general theory of relativity \cite{LL2}.     

\section{Statement of the problem}   

Our starting point is the system of equations of Newtonian hydrodynamics and gravity for an ideal liquid described by the local position- and time-dependent mass density $\rho(\textbf{r},t)$ and velocity $\textbf{v}(\textbf{r},t)$ fields  \cite{Milne,Bonnor,ZN}, which are related by the continuity equation
\begin{equation}
\label{continuity}
\frac{\partial \rho}{\partial t}+\nabla \cdot(\rho\textbf{v})=0.
\end{equation}
The equation of motion of the liquid is given by the Euler equation of hydrodynamics 
\begin{equation}
\label{2nd_law}
\frac{\partial \textbf{v}}{\partial t} +(\textbf{v}\cdot \nabla)\textbf{v} =-\nabla \phi
\end{equation}
where $\phi$ is the gravitational potential.  In Eq.(\ref{2nd_law}) the effects of the pressure are neglected which is a legitimate approximation in a variety of cosmological applications provided the velocities of the particles of the liquid are much smaller than the speed of light \cite{ZN}.  Additionally, a zero-pressure limit is a good approximation to the equation of state of dark matter.  The gravitational potential is determined by the density $\rho$  via the Poisson equation \cite{Milne,Bonnor}
\begin{equation}
\label{Poisson}
\nabla^{2}\phi=4\pi G(\rho-\rho_{0}),~~~\rho_{0}=\frac{\Lambda c^{2}}{4\pi G}
\end{equation}
where $G$ is the universal gravitational constant, and the characteristic density $\rho=\rho_{0}$ corresponding to Einstein's Universe is finite only in the presence of the cosmological constant $\Lambda$.   The density of Einstein's Universe corresponding to the current estimate of the cosmological constant (\ref{Lambda}) is  
\begin{equation}
\label{Einstein_density}
\rho_{0}\approx 10^{-29} g/cm^{3}
\end{equation}
We hasten to mention that the present day value of the average matter density is a small fraction of Einstein's density $\rho_{0}$ (\ref{Einstein_density}) again ruling out Einstein's static Universe.

According to the Euler equation (\ref{2nd_law}), the liquid is accelerated by the gradient of the gravitational potential $-\nabla \phi$. While the particles of the liquid attract each other, they are repelled by the uniform background due to the cosmological constant $\Lambda$.  The latter creates a possibility for a static solution $\rho=\rho_{0}$ which is, however, unstable \cite{Bonnor}.  

We also observe that if the right-hand side of the Euler equation (\ref{2nd_law}) would be multiplied by the negative of the electron charge to mass ratio, and $\phi$ identified with the electrostatic potential,  the system of equations (\ref{continuity})-(\ref{Poisson}) would describe (neglecting the effects of pressure) a non-relativistic electron plasma of the charge density $G\rho$ in the presence of a fixed compensating ion charge background of the density  $G\rho_{0}$ \cite{AL, breaking}.  Physics of this system is qualitatively different from that represented by Eqs.(\ref{continuity})-(\ref{Poisson}): electrons repel each other and are attracted by the ion background.  This admits a stable static solution $\rho=\rho_{0}$ corresponding to the state of local neutrality.      

\section{Gravitostatic waves} 

Let us consider one-dimensional motion along the $x$ axis, $\textbf{v}=(v,0,0)$, and seek solutions for the density $\rho$, velocity $v$ and gravitational potential $\phi$ that depend only on $\xi=x-ut$ where $u$ is the velocity of the wave.  Then Eqs.(\ref{continuity})-(\ref{Poisson}) transform into 
\begin{equation}
\label{wave_continuity}
(-u\rho+\rho v)'=0
\end{equation}
\begin{equation}
\label{wave_Euler}
-uv'+vv'=-\phi'
\end{equation}
\begin{equation}
\label{1d_Poisson}
\phi''=4\pi G(\rho-\rho_{0})
\end{equation}
where the prime is a shorthand for the derivative with respect to $\xi$.  

Solutions to this system of equations in purely Newtonian, $\rho_{0}=0$, cosmology called gravitostatic waves have been recently found  \cite{EBK}.  They exhibit universality:  with an appropriate choice of units the problem becomes parameter-free which means that all the solutions are qualitatively the same.  While the presence of the density scale $\rho_{0}$ in Eq.(\ref{1d_Poisson}) breaks universality, below we show that it admits solutions whose character changes with the average density $\overline{\rho}$, the central parameter of our investigation.  Specifically,  solitary wave solutions resembling those found in plasma physics \cite{breaking,EBKplasma} are found at $\overline{\rho}=\rho_{0}$, a possibility that cannot be realized if $\rho_{0}=0$.  
Integrating  Eq.(\ref{wave_continuity}) we find $-u\rho+\rho v=const=-u\overline{\rho}$ where the integration constant is fixed by the requirement of absence of average mass flux in the wave, $\overline{\rho v}=0$.  As a result one obtains a relationship
\begin{equation}
\label{integrated_wave_continuity}
\rho=\overline{\rho}\frac{u}{u-v}
\end{equation} 
that already appeared previously \cite{AL,breaking,EBKplasma,EBK}. It implies that underlying particles cannot travel faster than the wave, $v\leqslant u$, and that their velocity changes sign at $\rho=\overline{\rho}$:  the particles are moving in the positive $\xi$-direction in the region where $\rho>\overline{\rho}$ and in the negative direction if $\rho<\overline{\rho}$.  

Integrating Eq.(\ref{wave_Euler}) we arrive at the Bernoulli equation
\begin{equation}
\label{Bernoulli}
\frac{(u-v)^{2}}{2}+\phi=0
\end{equation}
where without loss of generality the integration constant is set to zero.  This constraints the gravitational potential to non-positive values, $\phi\leqslant 0$;  the upper limit $\phi=0$ is reached at points where the velocity of the particles equals that of the wave, $v=u$, and the density (\ref{integrated_wave_continuity}) is singular.   

Combining Eqs.(\ref{integrated_wave_continuity}) and (\ref{Bernoulli}) we find the dependence of the density on the potential
\begin{equation}
\label{density_vs_potential}
\rho(\phi)=\overline{\rho}\frac{u}{\sqrt{-2\phi}}
\end{equation}

\subsection{Mechanical analogy}

Similar to observations already made in previous studies \cite{breaking,EBK,EBKplasma}, if $\phi$ is viewed as a position of a fictitious particle of unit mass, $\xi$ as a time, and $4\pi G[\rho(\phi)-\rho_{0}]$ as a force,  Eq.(\ref{1d_Poisson}) parallels Newton's second law of motion for the particle in the field of the potential energy
\begin{equation}
\label{potential_energy}
U(\phi)=4\pi G\left (\overline{\rho}u\sqrt{-2\phi}+\rho_{0}\phi\right )
\end{equation}
Then the first integral of Eq.(\ref{1d_Poisson}) has the form
\begin{equation}
\label{1st_integral}
\frac{\phi'^{2}}{2}+U(\phi)=\frac{g^{2}}{2}
\end{equation}
where the integration constant $g^{2}/2$ is the energy of the fictitious particle.  The parameter $g$ is also the magnitude of the gravitational field $-\phi'$ at $\phi=0$ (to be definite we set $g>0$).  Choosing $\phi=0$ to be located at $\xi=0$, Eq.(\ref{1d_Poisson}) can be integrated as $\xi\rightarrow 0$ with the following asymptotic results for the potential,  density (\ref{density_vs_potential}) and velocity (\ref{Bernoulli})
\begin{equation}
\label{asymptotic}
\xi\rightarrow 0: ~~ \phi=-g|\xi|,~~~\rho=\frac{\overline{\rho}u}{\sqrt{2g|\xi|}},~~~ v=u-\sqrt{2g|\xi|}
\end{equation}
Exactly the same behavior was found previously in the problem without the cosmological constant \cite{EBK};  the density singularity at $\xi=0$ does not cause conceptual difficulties as it is integrable.  

The fact that the density diverges at the maximum of the gravitational potential can be explained with the help of the Bernoulli equation (\ref{Bernoulli}).   Indeed, in the reference frame of the wave, the original particles flow over the static potential energy landscape $\phi$ in the negative $\xi$ direction, and Eq.(\ref{Bernoulli}) is a statement of conservation of energy, an interpretation well-known in plasma physics \cite{breaking,EBKplasma}.  Then a particle starting at $\phi=-u^{2}/2$ with velocity $-u$ arrives at the maximum of the potential $\phi=0$ with zero velocity.  Thus, particle accumulation at the maximum of the potential is the reason why the density is singular.  This is essentially the mechanism responsible for the wave breaking effect \cite{breaking}.  The density singularity centered at the maximum of the gravitational potential is thus a caustic of the density field.  Previous analysis of the problem without the cosmological constant \cite{EBK} related this singularity to the so-called Zeldovich pancakes (ZPs) \cite{Z,ZS,web}, interim wall-like density singularities that appear at non-linear stages of development of the gravitational instability.  We contend that the same connection holds in the presence of the cosmological constant:  the singularity (\ref{asymptotic}) is a steady-state counterpart of the ZP possessing an infinite life time. 

Unlimited accumulation of particles at the maximum of the gravitational potential, an artifact of the zero-pressure approximation, is halted by finite pressure effects.  The reasoning originally given for the $\rho_{0}=0$ version of the problem \cite{EBK} carries over to the $\rho_{0}\neq 0$ case and will not be repeated here.

If $\rho_{0}=0$, the potential energy function (\ref{potential_energy}) is a monotonically decreasing function of $\phi$ constrained by the ``hard-wall" condition $\phi \leqslant 0$.  As a result the fictitious particle can only perform oscillatory motion which translates into already studied spatially periodic behavior of the potential, density and velocity \cite{EBK}.  

\begin{figure}
\includegraphics[width=1.0\columnwidth, keepaspectratio]{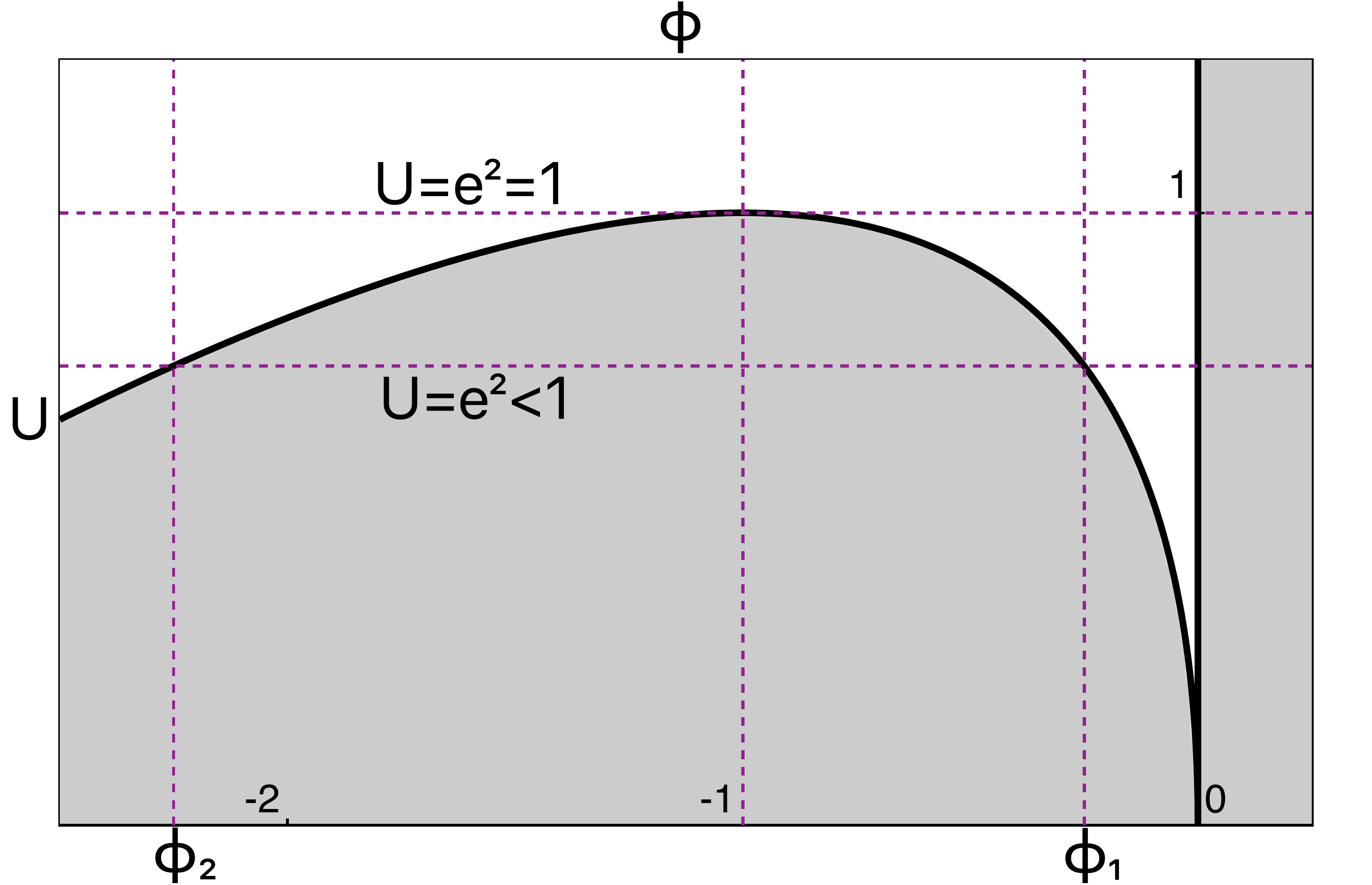} 
\caption{(Color online) Potential energy $U$ as a function of the gravitational potential $\phi$, Eq.(\ref{potential_energy}) in units of $U_{0}$ and $|\phi_{0}|$, Eq.(\ref{maximum}), respectively, or, equivalently, Eq.(\ref{energy_integral}), in dimensionless form.  The turning points of motion $\phi_{1,2}$ are given by Eq.(\ref{turning_points}).  For the energy and potential in the ranges $0\leqslant e^{2}\leqslant 1$ and $\phi_{1}\leqslant \phi \leqslant 0$ the motion is bounded.  Periodic solutions correspond to $0\leqslant e^{2} <1$ while the $e^{2}=1$ solution is a soliton.  Grayscale regions are not accessible for motion.}
\label{pwell}
\end{figure}

If, however, $\rho_{0}\neq 0$, the potential energy (\ref{potential_energy}) sketched in Figure \ref{pwell} has a maximum at
\begin{equation}
\label{maximum}
\phi=\phi_{0}=-\frac{\overline{\rho}^{2}u^{2}}{2\rho_{0}^{2}},~~~~U(\phi_{0})=U_{0}=\frac{2\pi G\overline{\rho}^{2}u^{2}}{\rho_{0}}
\end{equation}   
and the character of the motion of the fictitious particle depends on the dimensionless parameter
\begin{equation}
\label{ratio}
e^{2}=\frac{g^{2}}{2U_{0}}=\frac{\rho_{0}g^{2}}{4\pi G \overline{\rho}^{2}u^{2}}
\end{equation}
Like the amplitude of the gravitational field $g$ that first appeared in Eq.(\ref{1st_integral}), the quantity $e$ (assumed to be positive) is \textit{not} an independent parameter of the problem;  its dependence on $\overline{\rho}/\rho_{0}$ will be established below.

If $e^{2}>1$ the motion is infinite.  In terms of the original problem, this translates into unbounded variation of the gravitational potential which cannot be justified within the framework of the Newtonian gravity.  

If $e^{2}\leqslant 1$, then there are two turning points of motion
\begin{equation}
\label{turning_points}
\phi_{1,2}=-|\phi_{0}|\left (1\mp \sqrt{1-e^{2}}\right )^{2}
\end{equation}
corresponding to the zeros of the gravitational field which are the solutions of the equation $U(\phi)=g^{2}/2$.  If $\phi\leqslant \phi_{2}$ (the lower sign in Eq.(\ref{turning_points})), the motion is infinite and beyond the realm of Newtonian cosmology.  

If $\phi_{1}\leqslant \phi \leqslant 0$ the motion of the fictitious particle is finite which is relevant to the present study.  The ranges of variation of the potential and the density (\ref{density_vs_potential}) in the wave are thus given by   
\begin{equation}
\label{variation}
-\frac{\overline{\rho}^{2}u^{2}}{2\rho_{0}^{2}}\left (1-\sqrt{1-e^{2}}\right )^{2}\leqslant \phi \leqslant 0,\rho\geqslant\frac{\rho_{0}}{1-\sqrt{1-e^{2}}}
\end{equation}
We note that in the limit of the zero cosmological constant, $\rho_{0}\rightarrow 0$ ($e^{2}\rightarrow 0$), these inequalities reproduce their $\rho_{0}=0$ counterparts \cite{EBK}.   We also observe that since $\rho\geqslant \rho_{0}$, traveling wave solutions can only exist at an average density $\overline{\rho}$ larger than or equal to the density of Einstein's universe, $\overline{\rho}\geqslant \rho_{0}$. 

Integrating the first-order differential equation (\ref{1st_integral}) we find 
\begin{equation}
\label{general_solution}
\xi= \int_{0}^{\phi} \frac{d\phi}{\sqrt{g^{2}-2U(\phi)}}
\end{equation}
which implicitly gives a $\phi(\xi)$ dependence for $\xi\leqslant 0$;  the entire range of variation of $\xi$ is then included via a generalization $\phi(\xi)\rightarrow \phi(-|\xi|)$.

If $e^{2}<1$, the motion of the fictitious particle is oscillatory.  The gravitational potential $\phi$ is then a periodic function, $\phi(\xi+\lambda)=\phi(\xi)$, with the period $\lambda$ given by 
\begin{equation}
\label{period}
\lambda=2\int_{\phi_{1}}^{0}\frac{d\phi}{\sqrt{g^{2}-2U(\phi)}}.
\end{equation} 
The matter density $\rho(\xi)$ is also periodic with the same period.  The average density $\overline{\rho}$ that appeared in previous expressions can now be computed with the result 
\begin{equation}
\label{avdensity}
\overline{\rho}=\frac{2}{\lambda}\int_{-\lambda/2}^{0}\rho(\xi)d\xi=\rho_{0}+\frac{g}{2\pi G\lambda}
\end{equation}
where we employed the Poisson equation (\ref{1d_Poisson}) and the boundary condition $\phi'(\xi\rightarrow -0)=g$, Eq.(\ref{asymptotic}).  Eq.(\ref{avdensity}) relates the average density $\overline{\rho}$ and the amplitude of the gravitational field $g$ making it clear that only one of them is independent.  Since $1/\lambda$ has the meaning of a one-dimensional density of the ZP  singularities (\ref{asymptotic}), the expression for the average density (\ref{avdensity}) admits a natural physical interpretation:  the average density $\overline{\rho}$ is the sum of the background density $\rho_{0}$ of Einstein's Universe and contributions due to the ZP singularities, the second term of Eq.(\ref{avdensity}).  An individual ZP contributes $g/2\pi G$ into the areal matter density.     

\subsection{Soliton limit: generalities}    

Eq.(\ref{avdensity}) also implies that as $\overline{\rho}\rightarrow \rho_{0}$ from above or equivalently $e^{2}\rightarrow 1$ from below, the period $\lambda$ diverges.  This is the marginal case when a periodic solution turns into a solitary wave solution or soliton.  The magnitude of the gravitational field at the soliton center then follows by setting $e^{2}=1$ and $\overline{\rho}=\rho_{0}$ in the expression for the parameter $e^{2}$ in Eq.(\ref{ratio}): 
\begin{equation}
\label{soliton_field}
g=2u\sqrt{\pi G\rho_{0}}
\end{equation}
The potential and the density in the soliton, according to Eq.(\ref{variation}) then vary within the ranges $-u^{2}/2\leqslant \phi \leqslant 0$ and $\rho\geqslant \rho_{0}=\overline{\rho}$.  The latter along with Eq.(\ref{integrated_wave_continuity}) implies that all the particles in the soliton are moving in the direction of propagation of the wave, $v\geqslant 0$.  Their speed $|v|$ increases from zero at $\xi\rightarrow \pm\infty$ to $u$ at the soliton center $\xi=0$.  Since one soliton only carries along mass per unit area, it does not affect the bulk density.  That is why it is legitimate to characterize the soliton by the condition $\overline{\rho}=\rho_{0}$ even though $\rho\geqslant \rho_{0}$.  Likewise, the soliton solution satisfies the constraint of zero average mass flux, $\overline{\rho v}=0$, the condition behind Eq.(\ref{integrated_wave_continuity}), even though the particle velocity in the soliton does not change sign.   

As the average density $\overline{\rho}$ approaches the density of Einstein's Universe $\rho_{0}$, the period of the ZP lattice diverges according to
\begin{equation}
\label{soliton_limit}
\lambda(\overline{\rho}\rightarrow \rho_{0})=\sqrt{\frac{\rho_{0}}{\pi G}}\frac{u}{\overline{\rho}-\rho_{0}}
\end{equation}
and the wave may be viewed as a lattice of well-separated solitons.       

\subsection{Formulation in reduced units}

Further analysis is simplified if the velocity is measured in units of the velocity of the wave $u$, the density in units of the density of Einstein's Universe $\rho_{0}$, the gravitational potential $\phi$ in units of $|\phi_{0}|$, Eq.(\ref{maximum}), and the length in units of  
\begin{equation}
\label{length} 
l=\frac{\overline{\rho}u}{\sqrt{8\pi G\rho_{0}^{3}}},
\end{equation}
Then the velocity entering the Bernoulli equation (\ref{Bernoulli}) and the density (\ref{density_vs_potential}) are given in terms of the potential as
\begin{equation}
\label{velocity_density}
v=1-\overline{\rho}\sqrt{-\phi},~~~\rho=\frac{1}{\sqrt{-\phi}}
\end{equation} 
and Eqs.(\ref{potential_energy}), (\ref{1st_integral}) and (\ref{general_solution}) become
\begin{equation}
\label{energy_integral}
\frac{\phi'^{2}}{2}+U(\phi)=e^{2},~U(\phi)=2\sqrt{-\phi}+\phi
\end{equation}  
\begin{equation}
\label{quadrature_solution}
\sqrt{2}\xi=\int_{0}^{\phi}\frac{d\phi}{\sqrt{e^{2}-U(\phi)}}
\end{equation}
We now see that the parameter $e^{2}$ introduced in Eq.(\ref{ratio}) plays the role of dimensionless energy of the fictitious particle.

\subsubsection{Solitary wave solution}

\begin{figure}
\includegraphics[width=1.0\columnwidth, keepaspectratio]{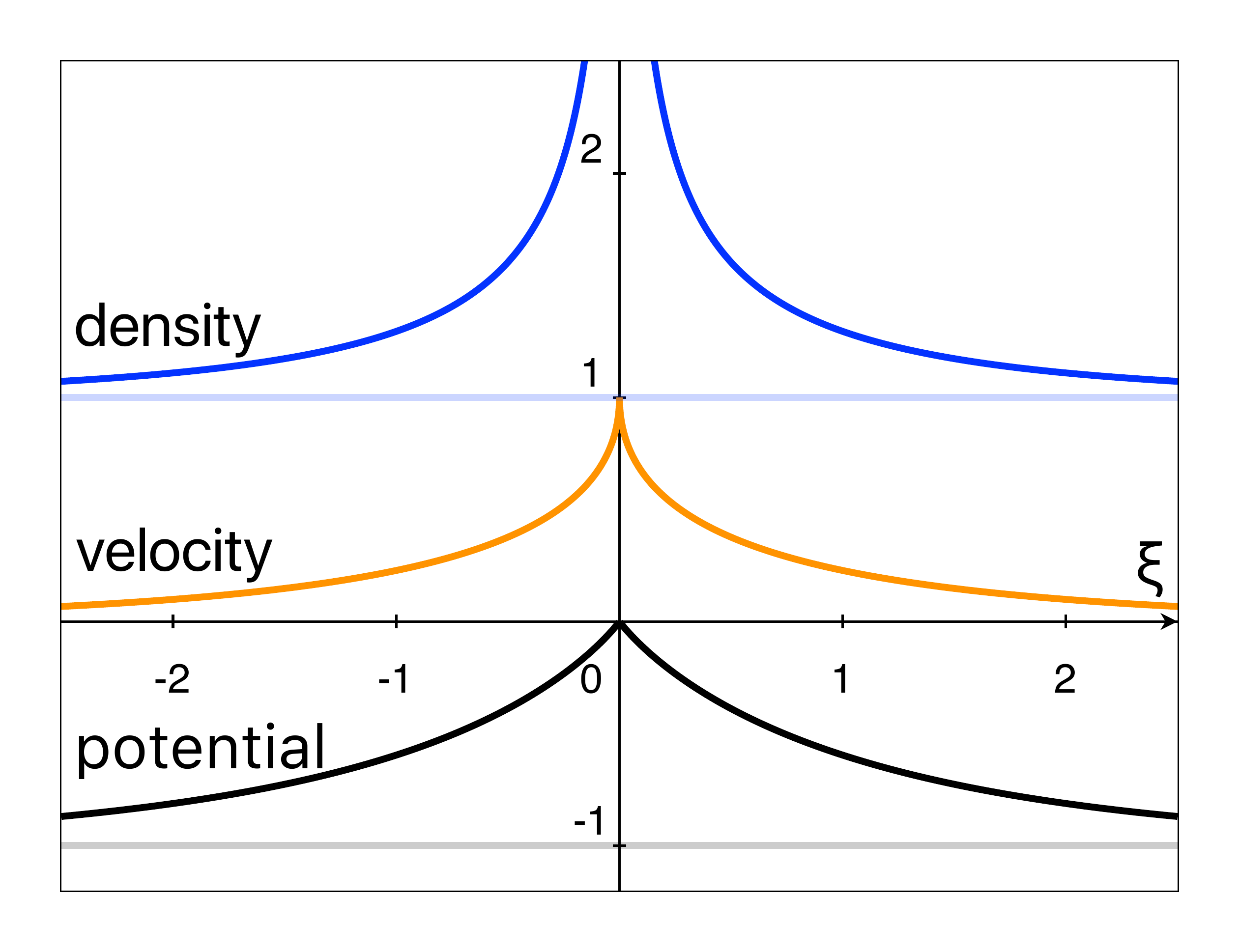} 
\caption{(Color online) Behavior of the matter density, Eq.(\ref{velocity_density}), velocity (\ref{velocity_modified}) and gravitational potential (\ref{soliton_potential}) in the ZP soliton propagating in the positive $\xi$ direction.  The units of the potential, velocity and density are $u^{2}/2$, $u$, and $\rho_{0}$, respectively.  The unit of length is $l_{0}$, Eq.(\ref{soliton_width}).  Faint horizontal black and blue lines are background levels of the potential and density, respectively.}
\label{soliton}
\end{figure}

Integrating Eq.(\ref{quadrature_solution}) for $e^{2}=1$ (the soliton limit) we find
\begin{equation}
\label{soliton_potential}
-\frac{|\xi|}{\sqrt{2}}=\sqrt{-\phi}+\ln(1-\sqrt{-\phi})
\end{equation} 
which together with the expressions for the velocity (evaluated at $\overline{\rho}=1$) and density, Eq. (\ref{velocity_density}), fully describes the soliton solution plotted in Figure \ref{soliton}. 
The unit of length here is the $\overline{\rho}=\rho_{0}$ case of Eq.(\ref{length}):
\begin{equation}
\label{soliton_width}
l_{0}=\frac{u}{\sqrt{8\pi G \rho_{0}}}
\end{equation}
While for $|\xi|\ll 1$, Eq.(\ref{soliton_potential}) matches the asymptotic result (\ref{asymptotic}) describing the soliton core, for $|\xi| \gg 1$, it becomes
\begin{equation}
\label{tail}
\phi(\xi)=-\left (1-e^{-|\xi|/\sqrt{2}}\right )^{2}
\end{equation}
which means that as $|\xi|\rightarrow \infty$ the asymptotic $\phi=-1$ limit is approached exponentially rapidly, and the length scale $l_{0}$ in Eq.(\ref{soliton_width}) also has a meaning of the soliton width.  We can now establish the range of validity of the soliton limit (\ref{soliton_limit}) as a condition of lack of overlap of the cores of neighboring solitons in the wave, $\lambda \gg l_{0}$.  This becomes $\rho_{0}\gg \overline{\rho}-\rho_{0}$, independent of the velocity of the wave $u$ or the gravitational constant $G$.    

\subsubsection{Periodic solution}
 
Integrating Eq.(\ref{quadrature_solution}) for general $e^{2}\leqslant1$ we find an expression    
\begin{eqnarray}
\label{general_potential}
-\frac{|\xi|}{\sqrt{2}}&=&e-\Big [\left (1-\sqrt{-\phi}\right )^{2}+e^{2}-1\Big ]^{1/2}\nonumber\\
&+&\ln\frac{1-\sqrt{-\phi}+\Big [\left (1-\sqrt{-\phi}\right )^{2}+e^{2}-1\Big ]^{1/2}}{1+e}
\end{eqnarray}
that for $e^{2}=1$ reduces to Eq.(\ref{soliton_potential}).  Evaluating the right-hand side of (\ref{general_potential}) at $\phi=\phi_{1}$, Eq.(\ref{turning_points}), the period of the wave can be determined as
\begin{equation}
\label{period_vs_e}
\lambda(e)=\sqrt{2}\left (\ln\frac{1+e}{1-e}-2e\right ).
\end{equation}
The solution (\ref{general_potential}) holds within the one full period of the wave $-\lambda/2\leqslant \xi \leqslant \lambda/2$, and needs to be periodically continued beyond that.  The $\lambda(e)$ dependence (\ref{period_vs_e}) is a monotonically increasing function that vanishes at $e=0$.   

In the soliton limit $e\rightarrow 1-0$ the period of the wave diverges in a logarithmic fashion, $\lambda\rightarrow -\sqrt{2}\ln(1-e)$.  This is equivalent to Eq. (\ref{soliton_limit}) but stated in terms of the difference $1-e$.  Remembering that the wavelength (\ref{period_vs_e}) is given in units of $l$ (\ref{length}) and employing the definition of $e^{2}$, Eq.(\ref{ratio}), the average density $\overline{\rho}$ can be expressed in terms of $e$ as
\begin{equation}
\label{density_vs_e}
\overline{\rho}=1+\frac{2\sqrt{2}e}{\lambda(e)}=1+\frac{2e}{\ln\frac{1+e}{1-e}-2e}
\end{equation}
This is a monotonically decreasing function that for $e\rightarrow 0$ diverges as $3/e^{2}$.  Combining with the definition of the parameter $e^{2}$ (\ref{ratio}) and restoring original physical units we find $g=2u\sqrt{3\pi G\overline{\rho}}$ which is the amplitude of the gravitational field in the $\rho_{0}=0$ gravitostatic wave \cite{EBK}.  In a similar fashion, an expression for the period of the $\rho_{0}=0$ wave, $\lambda=u\sqrt{3/\pi G \overline{\rho}}$ \cite{EBK}, can be recovered.

Eq.(\ref{density_vs_e}) can be also used to update the expression for the lower bound on the gravitational potential in the wave appearing in Eqs.(\ref{turning_points}) and (\ref{variation}) in the original physical units:
\begin{equation}
\label{updated_bound}
\phi_{1}=-\frac{u^{2}}{2}\left (1+\frac{2e}{\ln\frac{1+e}{1-e}-2e}\right )^{2}\left (1-\sqrt{1-e^{2}}\right )^{2}
\end{equation}
This is a monotonically increasing function of $e$:  as $e\rightarrow 0$ it tends to $-9u^{2}/8$ and reproduces the earlier $\rho_{0}=0$ result \cite{EBK} while as $e\rightarrow 1$ one finds $\phi_{1}=-u^{2}/2$ which represents the already analyzed soliton limit.

The $\phi_{1}(e)$ behavior as given by Eq.(\ref{updated_bound}) appears to be in contradiction with what is implied by Figure \ref{pwell}.  However the two cannot be directly compared because Eq.(\ref{updated_bound}) gives the lower bound on the gravitational potential in the original physical units while in Figure \ref{pwell} the potential is measured in units of $|\phi_{0}|$, Eq.(\ref{maximum}), which includes the $(\overline{\rho}/\rho_{0})^{2}$ ratio.  Similarly, the $\lambda(e)$ dependence (\ref{period_vs_e}) taken at its face value misrepresents the dependence of the measurable period of the wave on the parameter $e$ as $e\rightarrow 0$ which is equivalent to the $\rho_{0}=0$ case.  Indeed, according to Eq.(\ref{period_vs_e}) one finds $\lambda(0)=0$ while the physical period in this case is finite \cite{EBK} as was already explained following Eq.(\ref{density_vs_e}).  The reason behind this artifact is that the unit of length $l$, Eq.(\ref{length}), is singular as $\rho_{0}\rightarrow 0$.      

We see that while measuring the potential in units of $|\phi_{0}|$ (\ref{maximum}), the velocity in units of $u$, the density in units of $\rho_{0}$, and the length in units of $l$ (\ref{length}) simplifies the appearance of the already cumbersome general expression for the gravitational potential (\ref{general_potential}), these choices also create artifacts that are resolved only by reverting to the original physical units.  This can be avoided (at the price of complicating the appearance of Eq.(\ref{general_potential})) by choosing different units as follows:

The velocity, density, and gravitational potential will be measured in units of $u$, $\overline{\rho}$, and $u^{2}/2$, respectively, while the unit of the length will be based on the average density $\overline{\rho}$:
\begin{equation}
\label{av_length}
\overline{l}=\frac{u}{\sqrt{8\pi G \overline{\rho}}}.
\end{equation} 
This will have no affect on the $\rho(\phi)$ dependence while the $v(\phi)$ dependence in Eq.(\ref{velocity_density}) simplifies to
\begin{equation}
\label{velocity_modified}
v=1-\sqrt{-\phi},
\end{equation}
the expression for the period of the wave (\ref{period_vs_e}) acquires the form 
\begin{equation}
\label{new_period}
\lambda=\sqrt{2}\left (1+\frac{2e}{\ln\frac{1+e}{1-e}-2e}\right )^{3/2}\left (\ln\frac{1+e}{1-e}-2e\right ),
\end{equation}   
the expression for the average density (\ref{density_vs_e}) remains unaffected except that now it has to be viewed as a result for $1/\rho_{0}$.  Finally, Eq.(\ref{general_potential}) has to be modified according to the substitutions
\begin{equation}
\label{substitutions}
\xi\rightarrow \frac{\xi}{\overline{\rho}^{3/2}}, ~~~\phi\rightarrow \frac{\phi}{\overline{\rho}^{2}}
\end{equation}
where $\overline{\rho}$ is given by Eq.(\ref{density_vs_e}).

Eqs.(\ref{density_vs_e}) and (\ref{new_period}) now supply us with the $\lambda(\overline{\rho})$ dependence in the parametric form for arbitrary $0\leqslant e \leqslant 1$.  It is a monotonically decreasing function that diverges as $e\rightarrow 1$, the behavior already captured by the $\overline{\rho}\rightarrow \rho_{0}$ limit (\ref{soliton_limit}).  As $e\rightarrow 0$, the period approaches a finite value of $2\sqrt{6}$; this captures the behavior in the $\rho_{0}=0$ limit \cite{EBK}.  The implications of these conclusions in the original physical units is that the period of the gravitostatic wave is constrained by the inequality
\begin{equation}
\label{inequality}
\lambda\geqslant u\sqrt{\frac{3}{\pi G \overline{\rho}}}
\end{equation}       
where the lower bound is the $\rho_{0}=0$ result \cite{EBK}. 

\begin{figure}
\includegraphics[width=1.0\columnwidth, keepaspectratio]{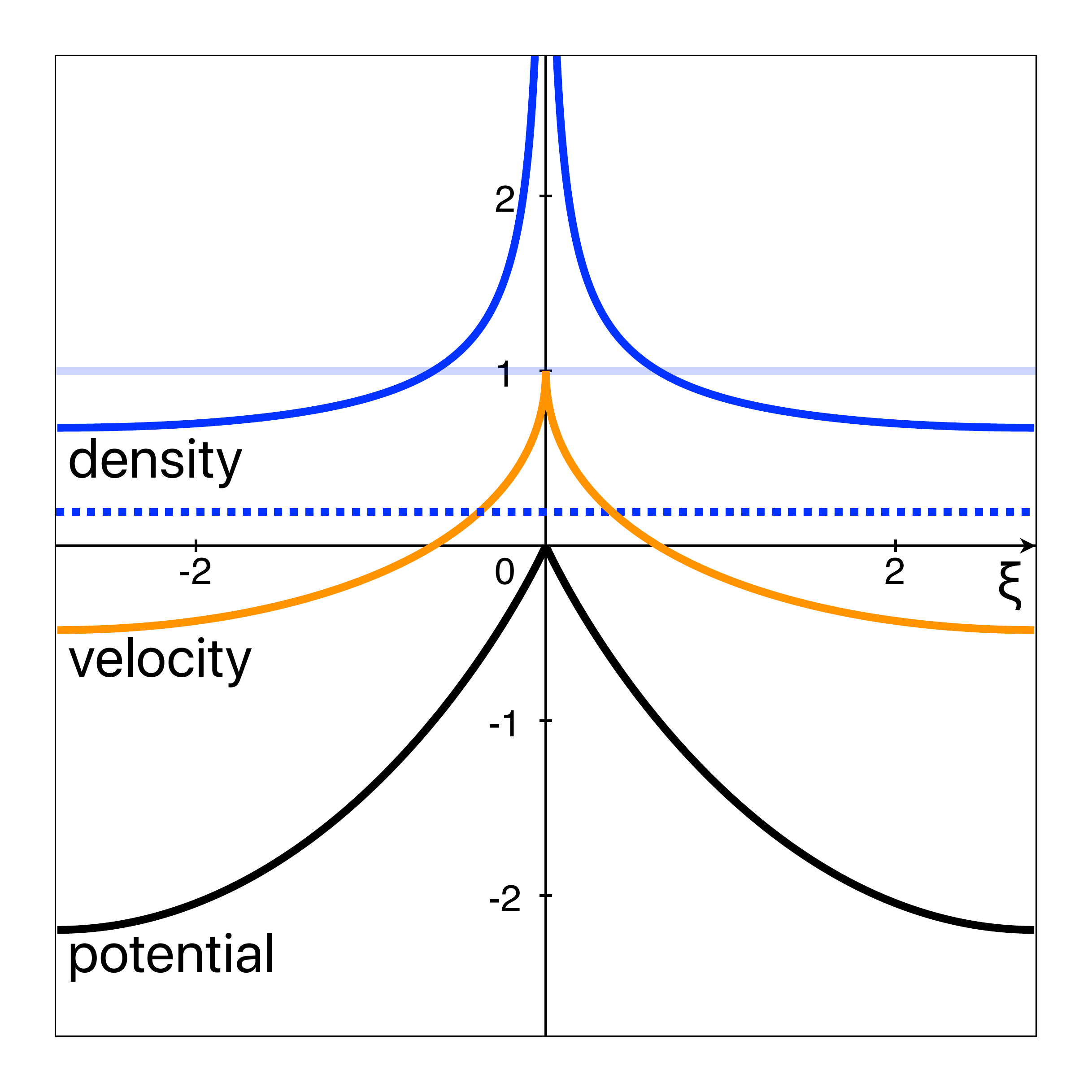} 
\caption{(Color online) Behavior of the matter density, Eq.(\ref{velocity_density}), velocity (\ref{velocity_modified}), and gravitational potential given by Eqs.(\ref{general_potential}) and (\ref{substitutions}), within one full period (\ref{new_period}) of the gravitostatic wave propagating in the positive $\xi$ direction.  The units of the potential, velocity and density are $u^{2}/2$, $u$, and $\overline{\rho}$, respectively.  The unit of length is $\overline{l}$, Eq.(\ref{av_length}).  The faint horizontal blue line is the average (unit) density while the dotted blue line is the background density given by $\overline{\rho}^{-1}$, Eq.(\ref{avdensity}).  The plot corresponds to $e=0.7$.}
\label{full_period}
\end{figure}

The typical behavior of the density, velocity and potential in the periodic gravitostatic wave propagating in the positive $\xi$ direction is  summarized in Figure \ref{full_period}.
The major qualitative difference from the soliton, Figure \ref{soliton}, can be seen in the behavior of the particle velocity which changes sign where the density matches the average density in the wave:  the particles in the high-density regions, $\rho>\overline{\rho}$ which include ZP singularities are traveling in the direction of the wave, while the particles in the low-density regions, $\rho<\overline{\rho}$ are traveling in the opposite direction.

\section{Steady flow solutions}

In an inertial reference frame traveling along with the wave the traveling wave solutions discussed in this work become static solutions corresponding to the presence of a fixed mass flux $j$ flowing in the positive $x$ direction.  The properties of these steady flow solutions can be inferred from already discussed properties of the traveling wave solutions via a substitution \cite{EBK}:
\begin{equation}
\label{steady}
u\rightarrow \frac{j}{\overline{\rho}}
\end{equation} 

\section{Conclusions}

To summarize, we have demonstrated that the introduction of the cosmological constant into the equations of Newtonian cosmology qualitatively changes the character of the traveling wave solutions.  In addition to periodic solutions that resemble previously identified gravitostatic waves \cite{EBK} we also found solitary wave solutions at an average matter density equal to that of Einstein's static Universe.  While Einstein's static Universe is known to be unstable with respect to small density fluctuations \cite{Eddington,Bonnor}, understanding outcomes of this instability is still the area of active scientific inquiry \cite{Z,ZS,web}.  It cannot be ruled out that at least some perturbations could lead to the final state that is a soliton traveling in the background Einstein's static Universe.  This is reasonable because the soliton is the steady-state counterpart of the ZP;  the latter are known to emerge at nonlinear states of gravitational instability \cite{Z,ZS,web}.    

We have also established that periodic traveling wave solutions can only exist at an average matter density exceeding that of Einstein's static Universe.  As the two densities approach each other, the period of the wave diverges, and in this limit the wave may be viewed as a train of well-separated solitons.  So as far as traveling wave solutions are concerned, the average density that is equal to that of Einstein's static Universe represents a marginal case that is akin to the point of phase transition.

\end{document}